\documentclass[twocolumn,conference,english]{IEEEtran}
\usepackage{algorithme}
\usepackage{graphics}
\usepackage{multicol}
\usepackage{xspace}
\usepackage{pstricks}
\usepackage{xcolor}
\usepackage{times}
\usepackage{amsthm}
\usepackage{epsfig}
\usepackage{fancyhdr}
\usepackage{epsf}
\usepackage{wrapfig}
\usepackage[ruled]{algorithm}
\usepackage{algpseudocode}
\usepackage{pst-node}
\usepackage{url}

\title{Low Complexity Link State Multipath Routing}

\author{ 
\IEEEauthorblockN{Pascal M\'erindol}
 \IEEEauthorblockA{
Universit\'e Catholique de Louvain (UCL)\\
Louvain la Neuve, Belgium\\
pascal.merindol@uclouvain.be} \and \IEEEauthorblockN{Jean-Jacques
Pansiot and St\'ephane Cateloin}

\IEEEauthorblockA{ LSIIT-CNRS,~
Universit\'e de Strasbourg (UdS)\\
Illkirch, France\\
\{pansiot,cateloin\}@unistra.fr} }

\begin{document}

\begin{figure}
\begin{center}
\rotatebox{90}{
\includegraphics[height=297mm]{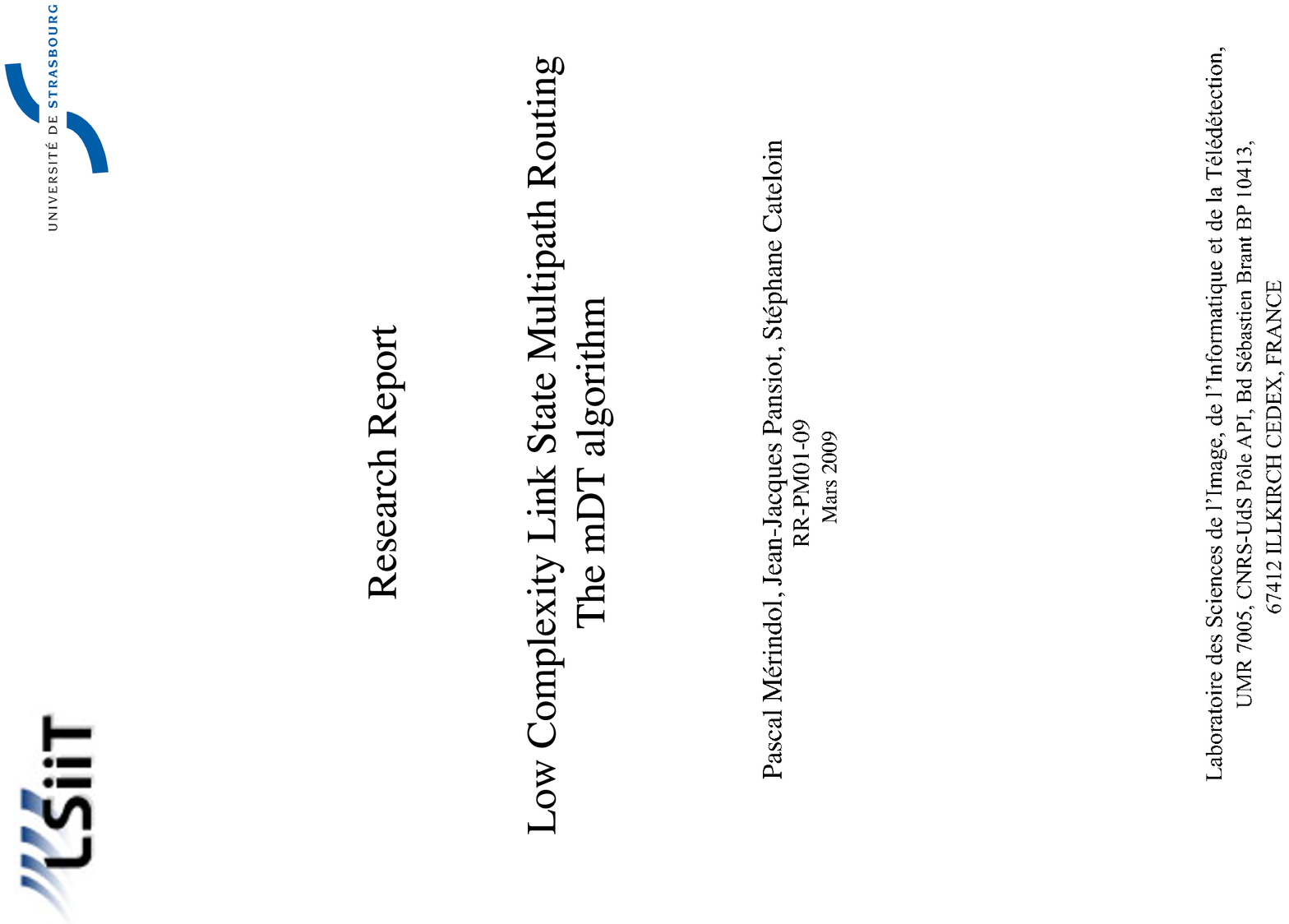}
}
\end{center}
\end{figure}

\maketitle
\begin{abstract}
%The reliability of an IP network against failures and congestions
%depends on the reaction time necessary for the convergence of the
%underlying routing protocol.
Link state routing protocols such as OSPF or IS-IS currently use
only best paths to forward IP packets throughout a domain. The optimality of sub-paths ensures consistency of hop by
hop forwarding although paths, calculated using Dijkstra's algorithm,
are recursively composed. According to the link metric, the diversity of
existing paths can be underestimated using only best paths. Hence,
it reduces potential benefits of multipath applications such as load
balancing and fast rerouting. In this paper, we
propose a low time complexity multipath computation algorithm able
to calculate at least two paths with a different first hop between
all pairs of nodes in the network if such next hops exist.
%it does not contain a
%bridge link.
%This approach allows
%to react faster to failures and congestions compared to the common
%unipath approach.
Using real and generated topologies, we evaluate and compare the complexity of our proposition with
several techniques. Simulation
results suggest that the path diversity achieved with our
proposition is approximatively the same that the one obtained
using consecutive Dijsktra computations, but with a lower
time complexity.
\end{abstract}

%\footnotetext{*Pascal M\'erindol is partly funded by Trilogy, a research project (ICT-216372) supported by the European Community. The views expressed here are those of the author only. The European Commission is not liable for any use that may be made of the information in this document.}
%\begin{IEEEkeywords}
%Path computation, Multipath routing, Next hops diversity.
%\end{IEEEkeywords}

\section{Introduction}
Routing is one of the key components of the Internet. Despite the potential benefits of multipath routing (e.g. \cite{mpathr} or \cite{anmpathr}), most
backbone networks still use unipath routing such as OSPF or IS-IS or their ECMP feature (Equal Cost MultiPath). With these routing protocols, the forwarding only changes upon
topology variations and not upon traffic variations. Dynamic multipath routing (e.g. \cite{COPE}, \cite{vuty}, \cite{WT} or \cite{apple2}) is
able to provide several services such as load balancing, to reduce delays and improve throughput, and fast rerouting schemes in case of failures.
The reliability of an IP network against failures and congestions
depends on the reaction time necessary for the convergence of the
underlying routing protocol. Proactive multiple paths calculation allows to accelerate this reaction time: pre-computed alternate paths can be directly used as backup paths without waiting for the routing protocol convergence.
This proactive mechanism can improve the network response in case of troubles where such backup paths exist. To provide these functionalities, the set of forwarding alternatives has to be large enough to achieve a good path diversity. 
However, current routers only support ECMP. This feature
corresponds to a simple variant of Dijkstra where equal cost paths
are inherited along the shortest path tree (SPT).
The optimality condition of sub-paths computed with ECMP restricts the number of loopfree paths and so reduces potential advantages of multipath routing.\\
In order to use multiple unequal cost paths between a pair of ingress and egress
routers, there are two forwarding possibilities. On the one hand, source multipath forwarding schemes can use
MPLS with a path signaling protocol (such as \textit{RSVP-TE} \cite{TE}) to establish any desired paths.
With this kind of approach, either the deployment is generalized in the whole
network and does not scale very well (proportional to the square of the number of routers), either
the reaction time can be as long as the notification delay on the
return path.\\ On the other hand, multipath routing protocols with hop
by hop forwarding needs to validate a set of next hops such that the recursive composition between neighbor routers does not create forwarding loops  (see \cite{OMPOSPF}, \cite{vuty} and \cite{PD}).
The first limitation is the complexity in time, space and the number of messages exchanged to compute and validate loopfree paths.\\
In this paper, we propose a simple hop by hop scheme that does not require a signaling protocol to validate loopfree paths. If the
validation procedure, whose goal is to verify the absence of loops,
is local (without exchanging any message) and does not involve all
routers, then the deployment can be incremental.
Our approach is equivalent to ECMP in terms of time, space and
message exchange complexity but allows to compute a greater diversity of forwarding alternatives.\\
%The goal is to keep the computation complexity as low as with the Dijkstra algorithm.\\
In this paper, we propose the following contributions:
\begin {enumerate}
%\item[-] a new graph decomposition analysis.
\item[-] a new graph decomposition analysis.
\item[-] two variants of the Dijkstra algorithm: Dijkstra-Transverse (DT) and multi-Dijkstra-Transverse (mDT).
\item[-] a proof that they compute at least two distinct next hops from the calculating node towards each node of the graph if such next hops exist.
\item[-] an evaluation of the efficiency and the complexity of our proposition compared to existing techniques.
\end {enumerate}
This paper is organized as follows. Section \ref{sec:related}
summarizes basic multipath routing notions and related work. Section
\ref{sec:dt} introduces our algorithms and their properties. Section
\ref{sec:results} presents our simulation results to underline the
relevance and the low time complexity of our proposition.
\section{Notations and context}\label{sec:related}
Table \ref{tab:tabnot1} lists the graph definitions used in the paper.
Notations are related to the multipath hop by hop forwarding context:
computed paths are loopfree and first hop distinct. We order paths
according to an additive metric $C$, and we focus on the best paths
having distinct first hops. To distinguish equal cost paths, we
consider the lexicographical order of first hops. For simplicity reasons we do not consider the multigraph issue: a
first hop is equivalent to a successor node, the
next hop. The valuation $w$ denotes the weight of each directed link
used by the routing protocol. Let us define a safety
property for distributed routing policies.
%\newline
\newtheorem*{definition}{Definition: Loopfree routing property at the router level}
\begin{definition}
A multipath routing protocol is loopfree if it always converges to a stable state
such that when any router
\textit{s} forwards a packet to any next hop \textit{v} towards any
destination \textit{d}, this packet never comes back to \textit{s}.
\end{definition}
\begin{center}
\begin{table}
\caption{\label{tab:tabnot1}Notations}
\begin{tabular}{|p{17mm}|*{1}{c|}}
\hline
Notations & Definitions\\
\hline \hline \bf $G(N, E, w)$ & oriented graph $G$ with a set of
nodes $N$, a set of \tabularnewline
               &edges $E$ and a strictly positive valuation of edges $w$.\\
%\hline
%\bf $|N|,|E|$ & respective cardinals of sets $N$ and $E$.\\
\hline \bf $e=(e.x,e.y)$ & edge $e \in E$ connecting node $x$ to
node $y$\tabularnewline
               & we assume that  $e^{-1}=(e.y,e.x) \in E$.\\
\hline
\bf $k^{-}(x)$, $k^{+}(x)$ & incoming and outgoing degrees of node $x$.\\
\hline
\bf $succ(x)$ & set of neighbors of node $x$ ($|succ(x)|=k^{+}(x)$).\\
\hline \bf $P_j(s,d) =$ & $j^{th}$ best loopfree path linking $s$ to $d$.
Recursively,\tabularnewline
               $(e_{1},..., e_{m})$&this is the best path whose first edge is distinct from\tabularnewline
               &the first edge of the $j-1$ best paths.\\
\hline \bf $C_j(s,d) =$ & cost of the path $P_j(s,d)$ \tabularnewline
        $\sum_{i=1}^{m}{w(e_{i})}$ & $1 \leq j \leq k^{+}(s),~0<m<|N|$.\\
\hline
\bf $NH_j(s,d)$ & $j^{th}$ best next hop computed on $s$ towards $d$. This is \tabularnewline & the first hop $e_{1}.y$ of $P_j(s,d)$.\\
\hline
\end{tabular}
\end{table}
\end{center}
With hop by hop link state multipath routing using multiple unequal
cost paths, two phases may be necessary to ensure loopfree routing: a
path computation algorithm and a validation process.
We do not consider validation processes using a signaling protocol (such as it can be done with distance vector routing messages, see \cite{vuty} for example).\\
With unipath or ECMP routing, the sub-path optimality condition guarantees the
correctness of next hop composition. To increase the number of valid
alternatives, the simplest rule to select a next hop $v$ on a router
$s$ (such that $v \in succ(s)$) is the \textit{downstream criteria} which can
be expressed as follows:
\begin{eqnarray}
C_1(v,d) < C_1(s,d) \label{eq:rule1}
\end{eqnarray}
This rule is referenced in the IS-IS  standard ISO 8473, is used in OSPF-OMP \cite{OMPOSPF} and is denoted LFI in
\cite{vuty} (with the particularity of avoiding routing loops even
in transient periods of topology changes). This rule is called
\textit{one hop vision} in \cite{PD} where Yang and Wetherall
introduce a set of rules whose flexibility allows to
increase the number of valid neighbors thanks to a \textit{two hops
vision}. This set of rules is more complex: the forwarding mechanism
is specific to the incoming interface and allows forwarding loops at the router
level but not at the link level. Thus, a packet is never forwarded through the same link but it can enter the same router twice.\\
Authors suggests that minimizing the queue level
should be the primary goal, however delays can increase if paths
contain several times the same router and this unnecessarily
consumes more resources (routers CPU, links bandwidth,...). We
consider that the queue usage is not the only resource to save.\\
In order to perform loopfree routing, the validation process needs
to compute a set of candidate next hops. A candidate next hop is a first hop of a computed path which is not yet validated for loopfree routing. On a given calculating
node (a root node $s$), the simplest way to obtain an exhaustive
candidate set is to compute the SPT of all
neighbor nodes. Thus, router $s$ can use the best costs information
of its neighborhood. This approach is denoted \textbf{kD} in the
following, and our analysis uses this technique as a reference. The
complexity of kD depends on the number of neighbors: $k^+(s)+1$
instances of the Dijkstra algorithm are necessary to compute the local and neighborhood best costs. If a router has a large number of interfaces, the computation time can be too long. Even if this calculation is typically done offline, when a congestion or a failure occurs during this period, the router is unable to perform the traffic switching.\\
Another way is to use an enhanced SPT
algorithm to locally compute multiple paths for each destination.
For example, algorithms and implementations presented in
\cite{pascoal} are designed to compute the set of $K$-shortest
loopfree paths, but do not guarantee that these paths are first hop
distinct. The $K$-shortest loopfree paths problem is not suited for
simple hop by hop forwarding. Indeed, in order to forward packets via these $K$ explicit paths,
a signaling protocol is necessary to mark routes from the ingress
router towards each egress router. Here we focus on distinct first
hops computation ($K\leq k^+(s)$), and paths are implicity stored as
candidate next hops. The objective of our approach is to compute a set of loopfree first hop disjoint paths with a lower complexity than \textbf{kD}.
%To accelerate the candidate next hops computation and provide
%a minimal guarantee in terms of next hops diversity, 
For this purpose, we calculate a set of costs $\{C_j(s,d)\}~{}_{\forall d \in N}$ containing at least
two entries for each destination node $d$ in the graph.
With an enhanced SPT algorithm able to compute such a set, rule
(\ref{eq:rule1}) becomes:
\begin{eqnarray}
C_j(s,d) - w(s,v) < C_1(s,d) \label{eq:rule2}
\end{eqnarray}
If $v=NH_j(s,d)$ satisfies rule
(\ref{eq:rule2}), then $(s,v)$ is a valid next hop.
Thus, the $j^{th}$ next hop $v$ can be used by $s$ to reach $d$ and
it satisfies the loopfree routing property at the router level. Note that: $\forall d \in N, C_j(s,d) - w(s,v) \ge C_1(v,d)$.\\
%This next hop can also be validated for rerouting issue similarly to
%loop free alternates (LFA) presented in \cite{LFA}. In this case the
%LFA criteria, $C_1(v,d) - C_1(v,s) < C_1(s,d)$, becomes $C_j(s,d) -
%C_1(s,v) - A^\star < C_1(s,d)$. This rule simply means that $v$ does
%not use $s$ as a its best next hop to reach $d$. Note that
%$A^\star=C_1(s,v)$ if link valuation is symmetric or
%$A^\star=w(v,s)$ if not ($C_1(v,s)\leq w(v,s)$). However, with this
%less strict rule, the alternate path via $v$ can only be used if
%link $(s,NH_1(s,d))$ fails and if there is no failure on the link
%$(v,NH_1(v,d))$. Hence, more alternate next hops can be validated
%but their usage is more limited. For example, it does not suit for
%load balancing purposes. This specific rerouting validation process
%also needs either consecutive instances of Dijkstra computation or
%an enhanced SPT. In the draft of LFA or UTURN (an extension of LFA,
%see \cite{UTURN}), authors suggest the use of the \textbf{kD}
%computation.
%Rerouting issues are not in the scope of this paper.
\begin{table}
\begin{center}
\caption{ \label{tab:tabnot2} Multipath terminology}
\begin{tabular}{|p{21mm}|*{1}{c|}}
\hline
  Terms    & Definitions  \\
\hline
\hline
\bf \textbf{ $branch_h(s)$} &\small{subtree of the SPT rooted at a neighbor $h$ of $s$} \\
\hline
\bf transverse edge &\small{an edge is transverse if it connects} \tabularnewline & \small{two distinct branches $branch_h(s)$ and} \tabularnewline & \small{ $branch_{h'}(s)$ or if it connects the root $s$}\tabularnewline & \small{and a node $n \neq h$ in a $branch_h(s)$}\\
\hline
\bf internal edge &\small{an edge $e$ is internal if it connects two nodes} \tabularnewline & \small{$e.x$ and $e.y$ belonging to a given $branch_h(s)$ }\tabularnewline &\small{and such that $e \notin branch_h(s)$}\\
\hline
\bf \textit{k}-\textit{transverse} path &\small{a path is k-transverse if it contains exactly}  \tabularnewline
& \small{\textit{k} transverse edges and no internal edge}\\
\hline
\bf Simple  &\small{a $1$-transverse path $(e_{1},...,e_{m})$} \tabularnewline \textbf{\textit{transverse} path} & \small{such that $(e_{1},...,e_{m-1})=P_1(s,e_{m-1}.y)$} \tabularnewline $\mathcal{P} \in Pt(s,d)$ & \small{and $e_{m}$ is a transverse edge ($e_m.y=d$)}\\
\hline
\bf Backward  &\small{a $1$-transverse path $(e_{1},...,e_{m})$ such that for} \tabularnewline \textbf{\textit{transverse} path} &\small{a $z$ ${}_{(1\leq z<m)}$, $(e_{1},...,e_{z}) \in Pt(s,e_{z}.y)$}  \tabularnewline  $\mathcal{P} \in Pbt(s,d)$ & \small{and $(e_{m}^{-1},...,e_{z+1}^{-1})=P_1(d,e_{z+1}.y)$}\\
\hline
\bf Forward  &\small{a $1$-transverse path $(e_{1},...,e_{m})$ such that for} \tabularnewline  \textbf{\textit{transverse} path} &\small{a $z$, $(e_{1},...,e_{z}) \in Pt(s,e_{z}.y)  \vee  Pbt(s,e_{z}.y) $ } \tabularnewline $\mathcal{P} \in Pft(s,d)$ & \small{and $(e_{z+1},...,e_{m})=P_1(e_{z+1}.x,d)$}\\
\hline
\end{tabular}
\end{center}
\end{table}
To sum up, our approach follows these three steps: \begin{enumerate}
\item it uses an unmodified link state
routing protocol such as OSPF or IS-IS to obtain topological
information,
\item it uses a multipath computation algorithm (see section \ref{sec:dt}) instead of Dijkstra to
compute candidate next hops,
\item it uses condition (2) to select valid next hops.
\end{enumerate}
\section{Candidate next hops computation}\label{sec:dt}
This section describes our path computation algorithms and an original edge partition analysis.  Given a root node $s$, the set
of edges of a graph can be partitioned into four subsets
(we consider both directions of each edge):
\begin{itemize}
\item [-] Edges corresponding to first hops  of primary paths.
\item [-] Edges belonging to sub-trees corresponding to \textit{branches}. %and edges in the opposite direction.
\item [-] \textit{Transverse edges} connecting two distinct branches or connecting the root $s$ and a branch without being the first hop of a primary path.
\item [-] \textit{Internal edges} linking nodes of the same branch without belonging to this branch.
\end{itemize}
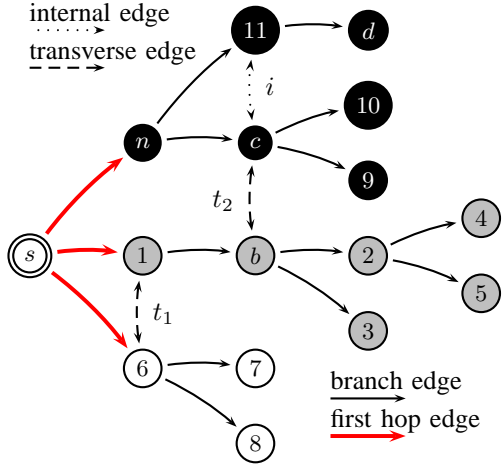
\begin{figure}
\begin{center}
\begin{picture}(150,165)
\cnodeput[doubleline=true] (0,2.5) {A} {\small{$s$}}
\cnodeput[fillstyle=solid,fillcolor=black] (4.5,5.5) {B}{\textcolor{white}{\small{$d$}}}
\cnodeput[fillstyle=solid,fillcolor=black] (3,4) {P}{\textcolor{white}{\small{$c$}}}
\cnodeput[fillstyle=solid,fillcolor=lightgray] (3,2.5) {C}
{\small{$b$}} \cnodeput[fillstyle=solid,fillcolor=black] (1.5,4) {D}
{\textcolor{white}{\small{$n$}}}

\cnodeput[fillstyle=solid,fillcolor=lightgray] (1.5,2.5) {E}
{\small{$1$}} \cnodeput[fillstyle=solid,fillcolor=lightgray]
(4.5,2.5) {F} {\small{$2$}}
\cnodeput[fillstyle=solid,fillcolor=lightgray] (4.5,1.5) {G}
{\small{$3$}} \cnodeput[fillstyle=solid,fillcolor=lightgray] (6,3)
{H} {\small{$4$}} \cnodeput[fillstyle=solid,fillcolor=lightgray]
(6,2) {I} {\small{$5$}} \cnodeput (1.5,1) {J} {\small{$6$}}
\cnodeput (3,1) {K} {\small{$7$}} \cnodeput (3,0) {L} {\small{$8$}}
\cnodeput[fillstyle=solid,fillcolor=black] (4.5,3.5) {M}
{\textcolor{white}{\small{$9$}}}
\cnodeput[fillstyle=solid,fillcolor=black] (4.5,4.5) {N}
{\textcolor{white}{\small{$10$}}}
\cnodeput[fillstyle=solid,fillcolor=black] (3,5.5) {O}
{\textcolor{white}{\small{$11$}}}

\psset{nodesep=2pt} \ncarc[linestyle=dotted]{<->}{P}{O}\trput{$i$}
\ncarc[linestyle=dashed]{<->}{J}{E}\trput{$t_1$}
\ncarc[linestyle=dashed]{<->}{C}{P}\tlput{$t_2$} \ncarc[linewidth=1.5pt,linecolor=red,linestyle=solid]{->}{A}{E}
\ncarc[linewidth=1.5pt,linecolor=red,linestyle=solid]{->}{A}{D} \ncarc[linewidth=1.5pt,linecolor=red,linestyle=solid]{->}{A}{J} \ncarc{->}{E}{C} \ncarc{->}{C}{F}
\ncarc{->}{F}{H} \ncarc{->}{F}{I} \ncarc{->}{J}{K} \ncarc{->}{J}{L}
\ncarc{->}{D}{O} \ncarc{->}{D}{P} \ncarc{->}{O}{B} \ncarc{->}{P}{N}
\ncarc{->}{P}{M} \ncarc{->}{C}{G}

\psline[linestyle=dotted]{->}(0,5.5)(1,5.5)
\rput[l]{0}(0,5.7){internal edge}
\psline[linestyle=dashed]{->}(0,5)(1,5)
\rput[l]{0}(0,5.2){transverse edge}

\psline{->}(4,0.6)(5,0.6)
\rput[l]{0}(4,0.8){branch edge}
\psline[linewidth=1.5pt,linecolor=red,linestyle=solid]{->}(4,0.1)(5,0.1)
\rput[l]{0}(4,0.3){first hop edge}
\end{picture}
\end{center}
\caption{\label{fig:interne}Edge partition example}
\end{figure}
These four subsets exhaustively describe $E$ because the set of
branches contains all nodes (except the root node $s$) in the graph.
Fig. \ref{fig:interne} illustrates an edge partition on a simple graph (some nodes are identified with a letter to facilitate the reading of section \ref{sec:dtprop}). In this
graph (we consider $w$ as a constant function), there are three branches (black, gray and white nodes), two
\textit{transverse} edges (dashed arcs denoted $t_1$ and $t_2$) and one \textit{internal}
edge (dotted arc denoted $i$). Edges $(s,n)$, $(s,1)$ and $(s,6)$
correspond to the three first hops (red arcs) linking $s$ to the three
branches.\\ With multipath hop by hop routing, the \textit{primary} path denotes
the optimal path depending on a given metric and a lexicographic
order to rank equal cost paths. Thus, for a given pair $(s,d)$, an
\textit{alternate} path is a path whose first edge is distinct from the first
one of the primary path $P_1(s,d)$.  More generally, if the forwarding mechanism is distributed such as with hop by hop routing, then all alternate paths are first hop distinct. Table
\ref{tab:tabnot2} summarizes all definitions related to
transverse paths terminology.
The path $((s,1),(1,b),(b,c))$ is \textit{simple transverse} and
the path $((s,1),(1,b),(b,c),(c,n))$ is \textit{backward transverse}. Paths
$\mathcal{P}=((s,1),(1,b),(b,c),(c,n),(n,11),(11,d))$ and $\mathcal{P'}=((s,6),(6,1),(1,b))$ are both \textit{forward transverse}. However,
$\mathcal{P}$ contains a sub path $((s,1),(1,b),(b,c),(c,n)) \in Pbt(s,n)$ whereas $\mathcal{P'}$
contains a sub path $((s,6),(6,1)) \in Pt(s,1)$.
The path $((s,6),(6,1),(1,b),(b,c))$ is $2$-transverse.\\
%Note that this figure represents the same network illustrated in figure \ref{fig:branch} in a hierarchical perspective.\\
The routing information base cannot directly use the set of candidate
next hops corresponding to the first hops of $1$-\textit{transverse} path to perform forwarding, since routing loops may occur. Our approach needs a validation
mechanism to select valid next hops among candidate next hops in order to guarantee
the safety of forwarding. In this paper, we consider the rule (\ref{eq:rule2}) introduced in section \ref{sec:related} to
validate candidate next hops. Due to space limitations, we do not
discuss and evaluate rules allowing to use a higher route diversity (see
\cite{MER2}).
\subsection{DT and mDT algorithms}  In \cite{MER}, we have proposed and
described the Dijkstra-Transverse algorithm (\textbf{DT}).  Here, we focus on DT properties that we have not presented in \cite{MER} (see section \ref{sec:dtprop}) and on a DT improvement that we call multi-DT (\textbf{mDT}). However, the basics of DT and mDT are similar.\\
To sum up, DT and mDT compute a multipath cost matrix on a given root node
(denoted $s$ in the following). A multipath cost matrix contains an
overestimation of best costs for all ($|N|-1$) destinations and via all
possible ($k^+(s)$) neighbors of $s$. The goal of these algorithms is to calculate a set of candidate next hops corresponding to costs associated to each neighbor. The calculation consists in two main stages:
\begin{enumerate}
\item[1-] Compute the best path tree and \textit{transverse} edges.
\item[2-] Compute \textit{backward} and \textit{forward transverse} paths.
\end{enumerate}
At each iteration, our algorithms compute the best $1$-transverse paths depending on the first
hop. Without an optimized structure to implement the best costs
vector, the complexity of DT for each calculating node \textit{s} is
in the worst case: $$O(|N|^2 + |E| + |N| \times k^+(s)) = O(|N|^2)$$
DT adds a time complexity proportional to the outgoing degree of the
given root node $s$ compared to Dijkstra.
%Yet $k^+(s) \leq |N|$, thus terms $|E|$
%and $|N| \times k^+(s)$ are cover by $O(|N|^2)$.
With a Fibonacci heap \cite{fibo} to implement the best costs
vector\footnote{The minimum extraction has an unitary cost
whereas the minimum suppression has an amortized cost in $O(log_2(|N|))$. For simplicity reasons, evaluations results that we present in this paper only rely on array lists.}, it is possible to reduce the time complexity to:
$$O(|N|log_2|N| + |E| + |N| \times k^+(s))$$
The set of candidate next hops computed with DT does not always include
all next hops corresponding to equal best cost paths. mDT
(see algorithm \ref{alg:multiDT}) is able to solve this problem. With mDT, only
the first computation phase of DT is modified by using a
next hop matrix denoted $Tp$. This matrix represents the existence
of a next hop per neighbor for each destination. $Tp$
is updated at each edge exploration. Candidate next hops recording follows a
transitive rule: $Tp(k,y)\leftarrow Tp(k,x)$ with $y \in succ(x),~k \in
succ(s)$. Initially, if $x=s$ then $Tp(y,y)\leftarrow y$. With ECMP, the
update of $Tp$ is performed only if $Tc(x)+w(x,y) \leq Tc(y)$. We have
chosen to generalize this approach to improve the upper
bound on the cost of \textit{forward transverse} paths composed with a
\textit{backward transverse} path. This generalization increases the number of
validated next hops. Indeed, during the exploration of the set of
successors of node $x$, if node $y$ is not already marked, it
inherits all forwarding alternatives of $x$, including when $(x,y)$
is an internal edge. In this case, the next hop inheritance is not
restricted to branches as with DT: $y$ is not the \textit{son} of
$x$ on a primary path. mDT allows to use all forwarding alternatives
already computed towards $x$.
%Algorithm mDT computes all alternate paths with one internal edge and one transverse edge if .
This set of paths is not limited to $1$-transverse alternatives, it
can contain alternate paths with several internal or transverse
edges. The mDT computation is based on the order of node exploration
which depends on the rank of costs stored in $Tc$. With mDT, the
first computation phase is able to calculate all candidate next hops corresponding to ECMP alternatives.
Recursively, the cost inheritance takes into account all the sets of
equal best cost paths for all marked nodes.
The complexity of mDT is slightly greater than the one of DT: for
each iteration of the main loop, $k^+(s)$ operations are necessary to execute the
inheritance of next hops and their costs. The worst case complexity
of mDT is in $O(|N|^2+E\times k^+(s))$ without an optimized structure
for $Tc$.
\begin{algorithm}
\caption{multi-Dijkstra-Transverse algorithm}\label{alg:multiDT}
\begin{algorithmic}[1]
\Procedure{multi-DT~}{$G(N, E, w), s$}
 \State $Mc_{k^+(s),|N|-1}$: Cost matrix
 \State $Tp_{k^+(s),|N|-1}$: Next hop matrix
\State $Tc_{|N|-1}$: List of best costs \State
$F_{|N|-1}$: List of father nodes \State $T_{|N|-1}$:
List of marked nodes \State $Mc(k,d)$,$Tp(k,d)$ and $Tc(d)$
$ \leftarrow \infty$,~$\forall d \in N,~k \in succ(s)$ \State
$Tc(s) \leftarrow 0$  
  \newline
  \Comment{\textit{
SPT and transverse path computation}}

\While {$|T| < |N|$} 
\State Choose the node $x$ ($x \notin T$) of minimum cost $Tc(x)$
 \For{$y \in succ(x)$}
 \For{$k \in succ(s) | Tp(k,x) \neq \infty$}
 \State Update $Tp(k,y)$
\If{$Mc(Tp(k,x),x)+w(x,y) < Mc(Tp(k,y),y)$}
 \State Update $Mc(Tp(k,y),y)$
 \EndIf
  \EndFor
 \If{$Tc(x)+w(x,y) < Tc(y)$}
 \State Update $Tc(y)$,~ $F_s(y)=x$
 %\State Update $Mc(Tp(y),y)$
 \EndIf
 \EndFor
 \State Put $x$ in $T$
 \EndWhile
\newline
\Comment{\textit{Backward and forward composition}}
 \For{$i:|N|\rightarrow 1$}\For{$y \in succ(s)$}
\If{$Mc(y,T(i))+w(T(i),F(T(i)))<Mc(y,F(T(i)))$} \State
\small{ Update $Mc(y,F(T(i)))$} \EndIf \EndFor
\EndFor\For{$i:1\rightarrow |N|$} \For{$y \in
succ(s)$}\If{$Mc(y,F(T(i)))+w(F(T(i)),T(i))<Mc(y,T(i))$}
\State Update $Mc(y,T(i))$ \EndIf \EndFor \EndFor
 \State Return $Mc$
\EndProcedure
\end{algorithmic}
\end{algorithm}

%\small{

%\begin{algorithm}
%\caption{Backward and Forward composition (step 2)}\label{alg:retav}
%\begin{algorithmic}[1] \Procedure{Composition~2~}{$s, w, Mc,
%F_{|N|-1}, T_{|N|-1}$}

%\State \small{Return $Mc$}
%\EndProcedure
%\end{algorithmic}
%\end{algorithm}

%}

\subsection{Properties of DT and mDT }\label{sec:dtprop}
In this section, we prove the ability of our algorithms
to compute at least two candidate next hops between each pair of nodes in the
graph if such next hops exist.
\newtheorem*{theoremes}{Property 1}
\begin{theoremes}
DT computes all $1$-transverse paths, and mDT computes all paths computed with DT and all equal best cost paths.
\end{theoremes}

The proof of these properties relies on next hops inheritance performed by DT and mDT (for more details, see \cite{MER}).\\
%More generally, our algorithms are able to compute at least one
%alternate next hop between all pair of nodes if such an alternative
%exists.
%For simplicity reasons, we focus on DT properties,
%but since the set of paths calculated with DT is included in the set
%of paths computed with mDT, these properties remain true with mDT. 
Now, let us define a major property of $1$-transverse paths.
\newtheorem*{theoreme}{Property 2}
\begin{theoreme}
If there exists an alternate path $P(s,d)$, then there exists a $1$-transverse
path between $s$ and $d$.
\end{theoreme}
%The \textit{1-branch distance} set contains at least two paths,
%between a given root node and all nodes of the graph if there exists
%an alternate path.
%Intuitively, the part of the path containing
%\textit{internal} edges between $c$ and $d$ is replaced by a primary
%path in the reverse direction between $c$ and $n$, followed by
%a primary path between $n$ and $d$.\\
The demonstration of this property relies on two lemmas. %On one
%hand, a path with $k$ transverse edges path contains a part of a
%\textit{l-transverse} path with $l=k-1$, so that recursively, it
%contains also a part of a \textit{1-transverse} path whose cost is
%equal or greater. On the other hand, if there exists an alternate
%path with one or more \textit{internal} edges and one transverse
%edge, then there exists a \textit{1-transverse} path using the same
%first edge.

\newtheorem*{lemma1}{Lemma 1}
\begin{lemma1}
If there exists an alternate path $\mathcal{P}$ from $s$ to $d$ then
there exists a path from $s$ to $d$ whose cost is not greater than the one of $\mathcal{P}$ and containing only one transverse
edge.
\end{lemma1}
\begin{IEEEproof}[Proof of Lemma 1]
Let $\mathcal{P} =(e_1,e_2,...,e_i,...,e_m)$ be an alternate path
from $s$ to $d$ where $e_i = (x, y)$ is the last transverse edge of
$\mathcal{P}$ and consider $P_1(s,x)$ the shortest path from $s$ to
$x$. Then either $P_1(s,x)$ is empty because $x = s$ and $i=1$, or
$P_1(s,x)$ is a primary path which is not longer than $(e_1,e_2,
...,e_{i-1})$. Let $\circ$ be the operator representing the path
concatenation. In both cases, there exists a path $P'$
such that $\mathcal{P'}=P_1(s,x)\circ(e_i,...,e_m)$ is an alternate
path with only one transverse edge and which is not longer than
$\mathcal{P}$.
\end{IEEEproof}

%The recursive proof on this first lemma is intuitive: by definition,
%to connect two different branches, it is necessary to use a
%transverse edge. If the first hop of a $k$-transverse path ($k>1$)
%is not included in the $1$-branch distance set, then there exists an
%alternate path, whose cost is necessarily lower or equal to the cost
%of this $k$-transverse path. This alternate path starts with the
%first hop connecting the penultimate branch used by the
%$k$-transverse path. Let us denote $a$ the last node used by the
%$k$-transverse path on the penultimate branch to link a source $s$
%and a destination $d$. Then we can consider the $1$-transverse path
%$\mathcal{P}$ such that the first part of $\mathcal{P}$ is
%$P_1(s,a)$ and the second part uses the same links between $a$ and
%$d$ used with the $k$-transverse path: this part contains exactly
%one transverse edge. This observation implies the existence of at
%least one alternate next hop in the $1$-branch distance set.
Figure \ref{fig:interne} illustrates lemma 1. The $2$-transverse path $\mathcal{P}=((s,6),(6,1),(1,b),(b,c))$
between $s$ and $c$ via the neighbor node $6$ uses $branch_1(s)$ to
reach the transverse edge $(b,c)$. There exists an alternate simple
transverse path $\mathcal{P'}=P_1(s,b)\circ((b,c))$. Note that the
existence of a path $P$ with several transverse edges implies that DT
(and mDT) implicitly records a $1$-transverse path
$\mathcal{P'}$ in the cost matrix $Mc$ with a cost lower or equal to the cost of $P$.

\newtheorem*{lemma2}{Lemma 2}
\begin{lemma2}
If there exists an alternate path from $s$ to $d$ with one
transverse edge, then there exists a $1$-transverse path linking $s$
and $d$.
\end{lemma2}

\begin{IEEEproof}[Proof of Lemma 2]
Let $\mathcal{P} =(e_1,...,e_i,...,e_m)$ be such an alternate path where
$e_i=(b,c)$ is the unique transverse edge.
% If $\mathcal{P}$ does not
%contain any internal edge, then it is a $1$-transverse path. 
Without loss of generality we may assume that $P_1(s,b)=(e_1,...,e_{i-1})$
is a primary path (see lemma1) without any internal edge.
Note that $(e_1,...,e_i) \in Pt(s,c)$.
%Let assume that there exists an alternate path
%$\mathcal{P}=(e_1,...,e_m)$ linking $s=e_1.x$ and $d=e_m.y$ whereas
%there is no $1$-transverse path between $s$ and $d$. $\mathcal{P}$
%uses a first hop $e_1.y$ which is not considered in the
%\textit{1-branch distance} set. If we denote
%$P_1(s,d)=(a_1,...,a_l)$, the primary path linking \textit{s} and
%\textit{d}, we know by definition that $e_1 \neq a_1$. If $e_1$ is
%not included in a branch, it is a transverse edge. We also know that
%$branch_{e_1}(s)$ and $branch_{a_1}(s)$ sets have a empty
%intersection.
To characterize the differences between transverse paths, we use an
``ancestor function".
%the
%function $F^i_s: N \rightarrow N, {}_{~0 \leq i < |N|}$. This
% is such that $F^i_s(x)=n$ if nodes $n$ and $x$
%belong to the same branch, and if $C_1(n,x)=i$ where
%$w(e)=1,~\forall e \in E$. The node $n$ is called an ancestor at $i$
%hops of node $x$. This relation is dependent of the SPT rooted on
%$s$.
An ancestor $a$ of a node $x$ is a node such that there exists a
primary path $P_1(a,x)$ included in the SPT rooted at $s$. The
closest common ancestor $n$ of nodes $x$ and $y$ is an ancestor of
$x$ and $y$ such that for any common ancestor $a$ of $x$ and $y$,
$a$ is also an ancestor of $n$.\\ Let $n$ be the closest common
ancestor of nodes $c$ and $d$.
%\footnote{A node
%$n=F^i_s(a)=F^j_s(b), ~a,b \in N$, such that $\sum {i+j}$ is
%minimized, is the closest common ancestor of nodes $x$ and $n$.}
\begin{enumerate}
\item If $n=c$ then there exists a forward transverse path
linking $s$ and $d$: a simple transverse path between $s$ and
$c$ and a primary path between $c$ and $d$.
\item Else if $n=d$ then there exists a backward transverse path linking $s$ and
$d$: a simple transverse path between $s$ and $c$ and a path in
the reverse direction of the primary path between $d$ and $c$ \footnote{We assume that $e \in E \Rightarrow e^{-1} \in E$.}.
%In both cases, paths are included in the $1$-transverse paths
%sets.
\item Else if $n \neq c,n \neq d$, then $n$ is the node where the branch including $d$ and $c$ is subdivided into two
sub-branches, one containing $c$, the other containing $d$ 
\footnote{Note that we know that $C_1(s,c)> C_1(s,n)$
and $C_1(s,d)> C_1(s,n)$.}. 
In this case, there exists a forward transverse path
linking $s$ and $d$ which contains a backward
transverse path $\in Pbt(s,n)$ and a primary path $P_1(n,d)$. 
\end{enumerate} Thus, in each case, the existence of a
\textit{1-transverse} path allowing to reach $d$ is
verified.\end{IEEEproof}

Figure \ref{fig:interne} illustrates lemma 2. Although the
alternate path $((s,1),(1,b),(b,c),(c,11),(11,d))$ is
not $1$-transverse because it contains an internal edge $(c,11)$,
there exists a forward transverse path
$((s,1),(1,b),(b,c),(c,n),(n,11),(11,d))$. In this case, the internal edge $i$ is bypassed with a backward
composition followed by a forward composition.
It allows to compute the alternate next hop $1$ to reach $d$.\\
Thanks to the backward and forward composition, if there exists a
$1$-transverse path, then DT finds it. These two phases allow to use
edges of the SPT in both directions. Moreover, DT considers all
transverse edges because, as it is the case for the classical Dijkstra algorithm,
all edges must be explored in order to mark all nodes.
The difference is that DT implicitly stores longer or equal cost
paths in the cost matrix.

\newtheorem*{theoreme2}{Corollary 1}
\begin{theoreme2}
For any pair of nodes $(s,d)$, if there exists an alternate path from $s$ to $d$,
then DT and mDT allow $s$ to compute at least two candidate next hops towards $d$.
\end{theoreme2}
\newtheorem*{theoreme3}{Corollary 2}
\begin{theoreme3}
If the graph contains no bridge edge, then DT and mDT allow $s$ to compute at least
two candidate next hops between any node and any other node of the
graph.
\end{theoreme3}

For a given destination, the corollary 1 allows to conclude that the
number of candidate next hops is at least $2$ if there exists an
alternate path linking $s$ and $d$.
Corollary $2$ is more specific,
if the network is 2-edge connected, then corollary 1 can be extended
for all pairs of routers.
%\newtheorem*{lemma4}{Lemma 4}
%\begin{lemma4}
%If an alternate candidate next hop is found by DT, it is also found
%by mDT. Moreover, all alternate optimal paths are found by mDT.
%\end{lemma4}

\section{Evaluation}\label{sec:results}
We use the Network Simulator 2 (ns2, \cite{ns2}) to compare several
routing approaches. ECMP is already implemented within the link
state module of ns2. We have extended ns2 to implement DT, mDT, kD
and the downstream criteria, rule (2), in the routing module (see \cite{implem} to find the implementation).
\subsection{Topologies and simulations setup}
We present results obtained on three different kinds of topologies.
The first category of networks are real topologies with actual IGP
weights (for confidentiality, we approximate their size in Table
\ref{tab:tableau}).
%The link
%valuation used in our simulation is based on propagation delays and
%others operators parameters (see \cite{geant} and \cite{abilene}).
Topologies denoted ISP1 and ISP2 are commercial networks covering an
European country. ISP3 and ISP4 are Tier-1 ISP networks.
%The second category of topologies
%were chosen among a inferred set of maps (given in \cite{oim}).
%These network topologies have been obtained through the
%\textit{mrinfo} tool. For networks where native multicast routing is
%enabled and \textit{mrinfo} is not filtered, this tool gives precise
%maps of router interconnections (see \cite{JJP}). For simplicity, we
%assume symmetry in connectivity and weight assignment and we
%consider an uniform link valuation.
The second category of topologies were chosen among the Rocketfuel
inferred set of maps given in \cite{rocket}.\\
We have also used the Igen topology generator (\cite{IGEN}) in order
to obtain a set of evaluation topologies of various sizes. We have
generated $10$ topologies containing between $20$ and $200$ nodes using the
$K$-medoid parameter, the delay-triangulation heuristic and a
$2$-sprint pop design. The $K$ parameter that determine the number of
routers per cluster is chosen such that $K=\frac{|N|}{10}$, so that each cluster contains approximatively $10$ routers for each
generated topology. These parameters offer a great physical
diversity to measure the relevance of our proposition to achieve the
same level of diversity as computed with $kD$. The link valuation
used for this third category is the inverse of the link capacity. The mean degree, denoted $k$, is approximatively the same for each generated topology: $k \sim 4$.
These networks represent access backbones and contain two kinds of
links: $155$Mbps for access links and $10$Gbps for backbone links (so
that weights of links are respectively $64$ and $1$).
%For all topologies, we have used an additive metric.\\ Figures
%\ref{fig:div}, \ref{fig:comp} and \ref{fig:val} concern results
%obtained with generated topologies, whereas table in figure
%\ref{fig:tableau} summarizes the main characteristics and results of
%real and inferred networks.
%The next three
%subsections gives an illustration of comparison between different
%techniques on the Igen topologies.

\subsection{Results}

%\twocolumn
\subsection{Diversity results}
\begin{figure}
\begin{center}
\rotatebox{270}{
\includegraphics[height=80mm]{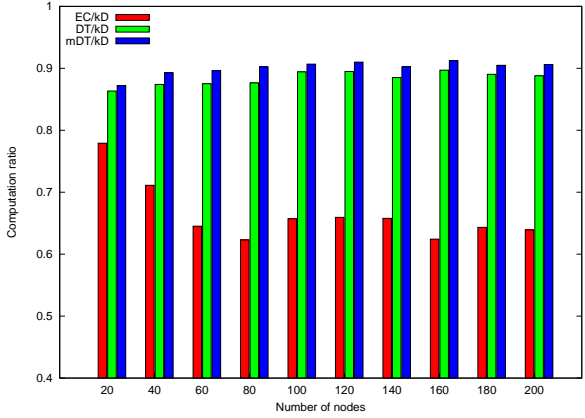}
}
\end{center}
\caption{\label{fig:div}Number of candidate next hops (Igen topologies)}
\end{figure}

First, we have measured the path diversity (see Fig.\ref{fig:div}). We have
calculated the total number of candidate next hops obtained with
ECMP (denoted EC), DT, mDT, and multiple Dijkstra computations
(kD). Results are represented as a performance ratio between
the considered technique and kD for all routers of a given network.
kD provides the best diversity but with a higher computation cost.
We observe that DT and mDT are able to compute approximatively $90\%$ of
candidate next hops obtained with kD, while ECMP obtains a
performance ratio only between $60\%$ and $80\%$.\\
\subsection{Complexity results}
\begin{figure}
\begin{center}
\rotatebox{270}{
\includegraphics[height=80mm]{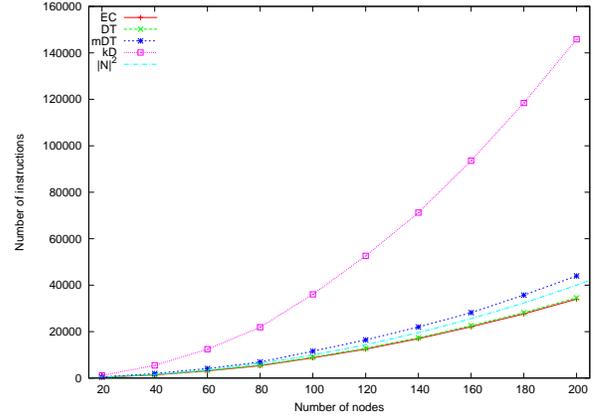}
}
\end{center}
\caption{\label{fig:comp}Number of operations (Igen topologies)}
\end{figure}

Then, we have compared the time complexities of the fore mentionned
algorithms (see Fig. \ref{fig:comp}). We have represented the
execution time measured in number of operations needed by DT, mDT and
kD to compute their set of candidate next hops. The number of operations is an average computed for each router. This value takes into account all operations necessary to extract the $min$ of $Tc$ and perform update of $Tc$, $Mc$ and $Tp$.
%\begin{figure*}
\begin{table*}
\begin{center}
\caption{Evaluation results on real and inferred topologies} \label{tab:tableau}
%\begin{multicols}{2}
%\end{multicols}
\small{
\begin{tabular}{||l||c|c||c|c|c|c||c|c|c|c||c|c|c|c||}
\hline
\multicolumn{1}{|c|}{} & \multicolumn{2}{c|}{} & \multicolumn{4}{c|}{Candidate next hops} & \multicolumn{4}{c|}{Validated next hops} & \multicolumn{4}{c|}{Number of operations}\\
\cline{4-15}
\multicolumn{1}{|c|}{Network} & \multicolumn{2}{c|}{Size} & mean & \multicolumn{3}{c|}{ratio/kD (\%)} & mean & \multicolumn{3}{c|}{ratio/kD (\%)} & mean & \multicolumn{3}{c|}{ratio/kD (\%)}\\
\cline{2-15}
\multicolumn{1}{|c|}{name} & $|N|$ & $|E|$ & \multicolumn{1}{c|}{ kD} & EC & DT & mDT  & \multicolumn{1}{c|}{kD} & EC & DT & mDT  & \multicolumn{1}{|c|}{kD} & EC & DT & mDT \\
\hline
\hline
\bf ISP1 & 25 & 50 & 1.46 & 76  & 97 & 97 & 1.10 & 97 & 100 & 100 & 489 & 60 & 66 & 75\\
\hline
\bf ISP2 & 50 & 200 & 3.58 & 43 & 93 & 97 & 1.79  & 69 & 89 & 94  & 6730 & 30 & 32 & 32.5 \\
\hline
\bf ISP3 & 110 & 350 & 2.70 & 55 & 89 & 92 & 1.45 & 82 & 97 & 99  & 8079  & 38 & 41 & 43.5 \\
\hline
\bf ISP4 & 210 & 880 & 3.73  & 44  & 86 & 88 & 1.81 & 72 & 96 & 99  & 41747 & 27 & 28 & 31 \\
\hline \hline
\bf Exodus & 79 & 294 & 3.58 & 44 & 88 & 96 & 1.73 & 58 & 94 & 99  & 5569 & 29 & 34 & 37\\
\hline
\bf Ebone & 87 & 322 & 3.49 & 46 & 90 & 96 & 1.76 & 77 & 93 & 99  & 9698 & 30 & 33 & 36\\
\hline
\bf Telstra & 104 & 304 & 2.30 & 72 & 92 & 95 & 1.30 & 90 & 98 & 99  & 6526 & 54 & 57 & 59\\
\hline
\bf Above & 141 & 748 & 5.29 & 34 & 86 & 97 & 2.50  & 58 & 89 & 99  & 40143 & 18.5 & 20 & 23\\
\hline
\bf Tiscali & 161 & 656 & 3.68 & 54 & 91 & 97 & 1.97 & 74 & 92 & 97  & 31044 & 27 & 29 & 32\\
\hline
\end{tabular}

}
\end{center}

%\end{table}

%\begin{multicols}{2}

%\end{multicols}
\end{table*}
We notice that the time saved with DT or mDT is really significant
compared to kD. The number of operations needed by kD is
approximatively $k\times|N|^2$ whereas mDT and DT
need approximatively $|N|^2$ operations. This complexity is
equivalent to the worst case of an ECMP computation. The time complexity upper bound is reached because some routers of Igen topologies
have a high degree of connectivity.\\
\begin{figure}
\begin{center}
\rotatebox{270}{
\includegraphics[height=80mm]{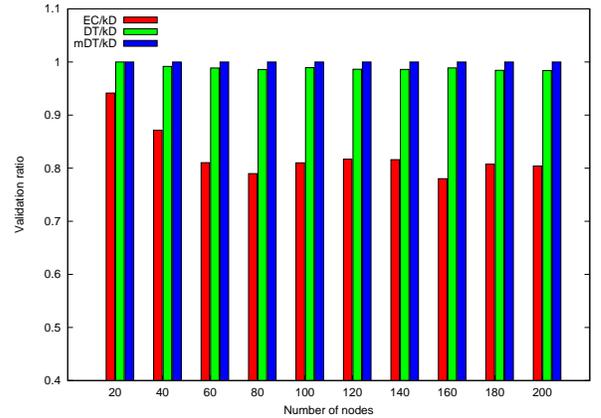}
}
\end{center}
\caption{\label{fig:val}Number of validated next hops (Igen topologies)}
\end{figure}
\subsection{Loopfree diversity results}
Finally, we have compared the number of validated next hops that are selected
with the downstream criteria (rule 2) depending on the computation
algorithm (see Fig. \ref{fig:val}). We remark that mDT allows to validate as many next hops as kD. This result can be
explained by the specific valuation function $w$ of our set of
generated topologies: there are only two very distant weights used
in these networks.
\subsection{General results and discussion}
Results given in Table \ref{tab:tableau} illustrate the same evaluation of
performance ratios and complexity on the set of real and inferred
topologies. For these sets of topologies, Table \ref{tab:tableau} also shows candidate and
valid next hops average per destination obtained with kD. Diversity ratio results
are similar to the ones obtained with Igen although degrees and
weights distributions are completely different. The main difference comes from
the time complexity evaluation. On these topologies, the maximum
degree of nodes is two times lower than with Igen topologies. The
measured complexity is far away from the theoretical worst case. More generally, several parameters, such as the valuation function $w$ or the degree distribution may strongly  influence complexity  measures, and thus the performance of algorithms. For example, if $w$ is a constant function, rule (2) is equivalent to ECMP. Thus, in this case, the number of valid next hops is the same for mDT, kD and ECMP. Another key point is the fact that the alternate paths which are not computed with mDT have a cost generally much more greater than the one of the primary path, that is why the ratio of loopfree alternatives between mDT and kD is close to $100\%$.\\ To summarize, although DT and mDT consume less processor resources than
kD, they are able to offer almost the same diversity in terms of
validated next hops.

\section{Conclusion}
Multipath routing enhances the network reachability and allows load
balancing to circumvent congestions or failures. However, the overhead imposed by signaling messages, the time and space complexity can hamper its
deployment. In this paper, we propose a simple scheme that is able to
generate a greater path diversity than ECMP with an equivalent overhead. Our path computation algorithms,
Dijkstra-Transverse, and its improvement multi-DT, allow to compute
at least two candidate next hops between all pairs of routers if such next hops exist. To validate candidate next hops in a
distributed manner, we have considered the simplest loopfree routing
rule, the downstream criteria. Our evaluations suggest that the gain
of time is very significant. We show that the number of next hops
validated with the downstream criteria is slightly the same using
mDT or a Dijkstra computation per neighbor. Moreover, our proposition
can be integrated in OSPF or IS-IS by replacing the path computation
algorithm without any change in the protocol. It can be
deployed incrementally, some routers using ECMP and others DT or
mDT. Our proposition can be extended to compute backup next hops only selected if a failure occurs.

\section*{Acknowledgement}
The research results presented herein have received support from Trilogy (\url{http://www.trilogy-project.eu}), a research project (ICT-216372) partially funded by the European Community under its Seventh Framework Programme. The views expressed here are those of the author(s) only. The European Commission is not liable for any use that may be made of the information in this document. The authors would like to gratefully acknowledge Pierre Francois and Olivier Bonaventure for their comments.

\bibliographystyle{IEEEtranS}
\bibliography{MPC-GIS09-techreport}

% Generated by IEEEtranS.bst, version: 1.13 (2008/09/30)
\begin{thebibliography}{10}
\providecommand{\url}[1]{#1}
\csname url@samestyle\endcsname
\providecommand{\newblock}{\relax}
\providecommand{\bibinfo}[2]{#2}
\providecommand{\BIBentrySTDinterwordspacing}{\spaceskip=0pt\relax}
\providecommand{\BIBentryALTinterwordstretchfactor}{4}
\providecommand{\BIBentryALTinterwordspacing}{\spaceskip=\fontdimen2\font plus
\BIBentryALTinterwordstretchfactor\fontdimen3\font minus
  \fontdimen4\font\relax}
\providecommand{\BIBforeignlanguage}[2]{{%
\expandafter\ifx\csname l@#1\endcsname\relax
\typeout{** WARNING: IEEEtranS.bst: No hyphenation pattern has been}%
\typeout{** loaded for the language `#1'. Using the pattern for}%
\typeout{** the default language instead.}%
\else
\language=\csname l@#1\endcsname
\fi
#2}}
\providecommand{\BIBdecl}{\relax}
\BIBdecl

\bibitem{implem}
``Implementation of dt and mdt in ns2,''
  \url{http://www-r2.u-strasbg.fr/~merindol/uploads/Research/DT.tar.gz}.

\bibitem{ns2}
``The network simulator- ns2,'' \url{http://www.isi.edu/nsnam/ns}.

\bibitem{apple2}
D.~Applegate and E.~Cohen, ``Making intra-domain routing robust to changing and
  uncertain traffic demands: understanding fundamental tradeoffs,'' in
  \emph{SIGCOMM}, 2003.

\bibitem{TE}
D.~Awduche, L.~Berger, D.~Gan, T.~Li, V.~Srinivasan, and G.~Swallow,
  ``{RSVP-TE} : Extensions to {RSVP} for lsp tunnels,'' RFC 3209, 2001.

\bibitem{mpathr}
R.~Banner and A.~Orda, ``Multipath routing algorithms for congestion
  minimization,'' \emph{IEEE/ACM Trans. Netw.}, 2007.

\bibitem{anmpathr}
I.~Cidon, R.~Rom, and Y.~Shavitt, ``Analysis of multi-path routing,''
  \emph{IEEE/ACM Trans. Netw.}, vol.~7, no.~6, 1999.

\bibitem{fibo}
T.~H. Cormen, C.~Stein, R.~L. Rivest, and C.~E. Leiserson, \emph{Introduction
  to Algorithms}.\hskip 1em plus 0.5em minus 0.4em\relax McGraw-Hill Higher
  Education, 2001.

\bibitem{WT}
S.~Kandula, D.~Katabi, B.~Davie, and A.~Charny, ``Walking the tightrope:
  Responsive yet stable traffic engineering,'' in \emph{SIGCOMM}, 2005.

\bibitem{rocket}
R.~Mahajan, N.~Spring, D.~Wetherall, and T.~Anderson, ``Inferring link weights
  using end-to-end measurements,'' in \emph{ACM SIGCOMM Internet Measurement
  Workshop}, 2002.

\bibitem{MER}
P.~M\'erindol, J.-J. Pansiot, and S.~Cateloin, ``Path computation for incoming
  interface multipath routing,'' in \emph{ECUMN}, 2007.

\bibitem{MER2}
P.~M\'{e}rindol, J.-J. Pansiot, and S.~Cateloin, ``Improving load balancing
  with multipath routing,'' in \emph{ICCCN}, 2008.

\bibitem{pascoal}
M.~Pascoal, ``Implementations and empirical comparison for k shortest loopless
  path algorithms,'' in \emph{The Ninth DIMACS Implementation Challenge: The
  Shortest Path Problem}, 2006.

\bibitem{IGEN}
B.~Quoitin, ``Topology generation through network design heuristics,''
  \url{http://www.info.ucl.ac.be/~bqu/igen/}.

\bibitem{OMPOSPF}
C.~Villamizar, ``Ospf optimized multipath (ospf-omp):
  draft-ietf-ospf-omp-02.txt,'' IETF, Draft, Feb. 1999.

\bibitem{vuty}
S.~N. Vutukury, ``Multipath routing mechanisms for traffic engineering and
  quality of service in the internet,'' Ph.D. dissertation, 2001.

\bibitem{COPE}
H.~Wang, H.~Xie, L.~Qiu, Y.~R. Yang, Y.~Zhang, and A.~Greenberg, ``Cope:
  traffic engineering in dynamic networks,'' in \emph{SIGCOMM}, 2006.

\bibitem{PD}
X.~Yang and D.~Wetherall, ``Source selectable path diversity via routing
  deflections,'' in \emph{SIGCOMM}, vol.~36, 2006.

\end{thebibliography}

\end{document}